\documentstyle[prd,preprint,aps,floats,epsfig]{revtex}  
\draft
\newcommand{\beq}{\begin{equation}}
\newcommand{\eeq}{\end{equation}}
\newcommand{\bea}{\begin{eqnarray}}
\newcommand{\eea}{\end{eqnarray}}

\begin{document}

\title{Proof of factorization for  electroproduction of multiple 
mesons and exclusive $\gamma^* \gamma$ production of multiple hadrons}

\author{Andreas Freund}

\address{I.N.F.N, Sezione Firenze, Lg. Enrico Fermi 2, 50125 Florence, Italy}

\maketitle

\begin{abstract}
In the following, we will present a generalization of the proof
of factorization for  electroproduction of vector mesons to the 
production of several mesons such as  $\pi^+\pi^-$ and a proof of 
factorization for exclusive $\gamma^* \gamma$ production of several 
hadrons for example $p \bar p$ or $\pi^+\pi^-$ to all orders 
in perturbation theory up to power suppressed terms.\newline
PACS: 12.38.Bx, 13.85.Fb, 13.85.Ni\newline
Keywords: Factorization, Parton Distributions, Distributional Amplitudes

\end{abstract}

\section{Introduction}
\label{intro}

In the last few years several proof's of factorization for various exclusive,
processes were given in \cite{cfs,rad,ca}, which contain novel nonperturbative
functions, the so-called skewed parton distributions. This note is designed 
to expand this still rather meager number of processes by 
generalizing the proof of exclusive,   electroproduction 
of vector mesons given in \cite{cfs} to an arbitrary number of mesons\cite{4} 
first discussed in \cite{brod,koep}, where however no all order proof was given. 
Furthermore, we will prove factorization to all 
orders in perturbation theory for exclusive $\gamma^*\gamma$ production 
of hadrons like $\pi^+\pi^-$ or $p\bar p$ \cite{4a}. All the above statements 
imply factorization of the amplitude of the processes under consideration up 
to power suppressed terms.

As will be shown, the generalization of the proof of Ref.\ \cite{cfs} is 
almost trivial except for a class of reduced diagrams which were considered
in Ref.\ \cite{cfs} but found not to contribute in 
$\rho$ and $J/\psi$ production due to C-parity. 
However they do contribute in, for example, $\pi^+\pi^-$ production 
which is now seriously being considered phenomenologically \cite{diehl,polya}. 
Furthermore,
in the case of multiple meson production, the meson distribution amplitude
in the factorization formula generalizes to a generalized distribution 
amplitude as first discussed in principle terms in \cite{koep} and in the
explicit form of a two pion distribution amplitude in Ref.\ \cite{diehl,polya}.
The most noteworthy and novel feature of this function is that it does not have to be
real valued, i.\ e.\ , can be complex.

Exclusive $\gamma^*\gamma$ production of hadrons \cite{Brod89,diehl} is interesting 
in its own right, since it enables one to expand the type of processes
which are sensitive to, in this case, generalized distribution amplitudes 
to the realm of $e^+e^-$ colliders. In principle, these type of reactions 
have already been measured but the experimental cuts employed will probably 
have excluded such exclusive events.

The paper is organized in the following way, first we will give the 
generalized proof for exclusive,   electroproduction of multiple 
mesons $\gamma_L^*(q) + P(p) \rightarrow \sum_{i=1}^{N}V(v_i) + P(p')$ and then 
the proof for exclusive $\gamma^*\gamma$ production of multiple hadrons 
$\gamma^*(q) + \gamma(q') \rightarrow \sum_{i=1}^{N}H(h_i)$ where the
letters in brackets denote the momenta of the particles and $H$ can also mean 
$\bar H$. Finally we will give concluding remarks.

\section{Proof of factorization for exclusive,  electroproduction 
of multiple mesons}
\label{genproof}

\subsection{Statement of Factorization Theorem}
\label{state}

The amplitude of exclusive, electroproduction of multiple mesons takes the 
following, factorized form
\bea
M =&\sum_{i,j}& \int^{1}_{0}dz\int^{1}_{-1+x}dx_1 P_{i/p}(x_1,x_1-x,\mu^2) 
H_{ij}(x_1/x,(x_1-x)/x,z,\mu^2)V_j(z,\zeta_1,...,\zeta_N,\mu^2)\nonumber\\
&+&\mbox{power suppressed terms}.
\label{faceq}
\eea

The steps in the proof of Eq.\ (\ref{faceq}) are given below, following 
Ref.\ \cite{libster,ster}:

\begin{itemize}

\item Establish the non-ultraviolet regions in the space of loop momenta
      contributing to the amplitude.

\item Establish and proof a power counting formula for these regions.

\item Determine the leading regions of the amplitude.

\item Define the necessary subtractions in the amplitude to avoid double
      counting.

\item Taylor expand the amplitude to obtain a factorized form.

\item Show that the part containing the long-distance information can be
      expressed through matrix elements of renormalized, bi-local, gauge
      invariant operators of twist-2.

\end{itemize}

We will use the conventions as well as the expressions for the particle 
momenta of Ref.\ \cite{cfs}. It is convenient to use light-cone coordinates 
with respect to the collision axis \cite{5},
the particle momenta then take the form
\begin{eqnarray}
   p^{\mu } &=& \left ( p^{+},\frac {m^{2}}{2p^{+}},{\bf 0}_{\perp } \right ),
\nonumber\\
   q^{\mu } &\simeq& \left ( -xp^{+},\frac {Q^{2}}{2xp^{+}},{\bf 0}_{\perp }
\right ),
\nonumber\\
   V^{\mu } &\simeq& \left (\frac{\Delta ^{2}_{\perp }+m^2_V}{Q^{2}},
      \frac {Q^{2}}{2xp^{+}},{\bf \Delta }_{\perp }
   \right ),
\nonumber\\
   \Delta ^{\mu } &\simeq& \left (xp^{+},
      -\frac {\Delta ^{2}_{\perp } + m^{2}x}{2(1-x)p^{+}},
      {\bf \Delta }_{\perp } \right ).
\label{momenta}
\end{eqnarray}
Here, $x$ is the Bjorken scaling variable, $Q^{2}$ is the virtuality of the 
initial photon, $m^{2}$ is the proton mass, $t=(p-p')^2 = \Delta ^{2}$ is
the momentum transfer squared, $V = \sum_{i=1}^N v_i$ is the sum of the
momenta of the produced mesons and
$\simeq$ means ``equality up to power suppressed terms''. We work in the
kinematic limit of 
$Q^2/m_{V}^2 \rightarrow \infty$ with $Q^2/W^2 = \mbox{const.}$.

Our proof is limited to a longitudinally 
polarized photon initiating the reaction. Photons with transverse 
polarization will not be considered here \cite{5a}.

\subsection{Regions}
\label{region}

The steps leading to the generalized reduced graph are identical to the 
steps 1-3 in Sec.\ IV of Ref.\ \cite{cfs}. Meaning that one first scales
all momenta by a factor $Q/m$,
secondly one uses the Coleman-Norton theorem to locate all pinch-singular 
surfaces in the space of loop momenta in the zero mass limit and finally 
identifies the relevant regions of integration as neighborhoods of 
these pinch-singular surfaces.

\begin{figure}
\centering 
\mbox{\epsfig{file=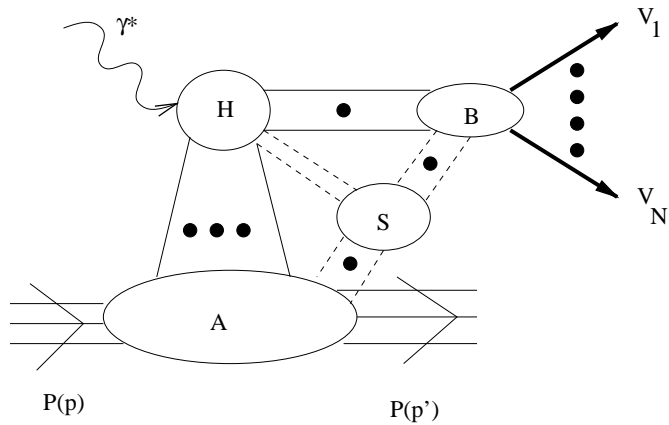,height=5.5cm}}
\vspace*{5mm}
\caption{Generalized reduced graph for exclusive, electroproduction of multiple
mesons.}
\label{gengraph}
\end{figure}

Due to the fact that one is only changing the final state produced by the
collinear-to-B graph, the generalized reduced graph has to be the same
as in Ref.\ \cite{cfs}, except for the final mesonic state and is given in 
Fig.\ \ref{gengraph}. Note that, H refers to the hard subgraph of the process, 
A to the subgraph which is collinear to the incoming and outgoing proton, B to 
the subgraph which is collinear to the produced mesons and S denotes the subgraph 
containing lines which only have small momentum components as compared to 
the large scale $Q$. The dots represent any number of additional partons 
connecting the different subgraphs or additional mesons in the final state.

\subsection{Power Counting}
\label{power}

For the same reason as mentioned in the previous subsection, the power counting
formula as established in Sec.\ VA of Ref.\ \cite{cfs} remains unaltered. The 
contribution to the amplitude from a neighborhood of a pinch-singular surface
$\pi$ behaves like $Q^{p(\pi)}$, modulo logarithms, in the large-$Q$ limit with
the power given by

\bea
p(\pi) &=& 3 - n(H) - \#(\mbox{quarks from S to A, B}) - 3\#(\mbox{quarks from 
S to H})\nonumber\\
& &- 2 \#(\mbox{gluons from S to H}).
\label{powerform}
\eea   
where $n(H)$ is the number of collinear quarks and transversely polarized
gluons attaching to the hard subgraph H.

\subsection{Leading regions}
\label{leadreg}

\begin{figure}
\centering
\mbox{\epsfig{file=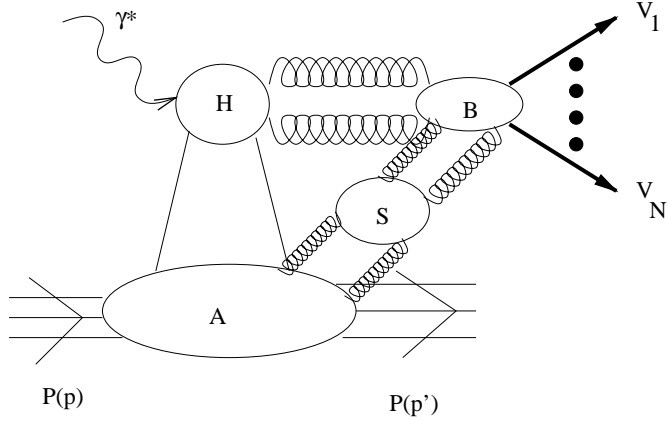,height=5.5cm}}
\caption{Additional graph contributing to the leading region of the process.}
\label{leadreg1}
\end{figure}

The leading regions are the same as the ones considered in Sec.\ VB of 
Ref.\ \cite{cfs}
 and yield a leading 
power for the amplitude of $Q^{-1}$ according to Eq.\ (\ref{powerform}). 
The discussion in Sec.\ VB of Ref.\ \cite{cfs} of the leading regions, 
as well as endpoint contributions, is unchanged since the arguments presented 
in \cite{cfs} are independent of the number of mesons produced. Note that, as
discussed below, the graph in Fig.\ \ref{leadreg1} has to be reconsidered.

This leading region Fig.\ \ref{leadreg1} is the same as Fig.\ 8c in 
Ref.\ \cite{cfs}, which was referred to as a glueball graph by the authors of 
Ref.\ \cite{cfs}. 
This graph did not contribute in their analysis since it is zero for $\rho$ and
$J/\psi$ production due to C-parity, but it does contribute
in our case to the leading power in $Q$. Since this type of graph is not an 
endpoint-contribution-type graph, we do not have to worry about it possibly 
ruining the factorization of the amplitude.

\begin{figure}
\centering 
\mbox{\epsfig{file=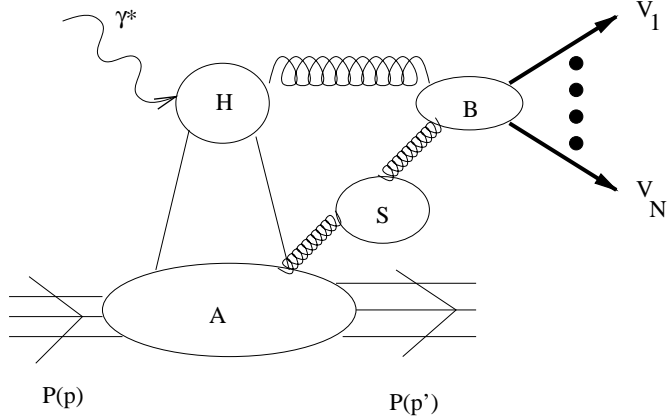,height=5.5cm}}
\caption{Another possible leading region.}
\label{dijet}
\end{figure}

Another possibility would be a graph of the type of Fig.\ \ref{dijet}. 
However, if the collinear-to-B gluon is a scalar gluon, i.e., has a polarization
vector proportional to its momentum ($k_{B}$), one obtains the expression 
$k_{B}\cdot H$ of the scalar gluon coupling to the hard part H. Now one can use
Ward identities as given in, for example, Ref.\ \cite{cfs,col} to show that the 
sum over gauge invariant sets of graphs where scalar gluons connect either A to
H or B to H, cancel. This is necessary since otherwise these type of graphs would
be leading in $Q$, as compared to Fig.\ \ref{leadreg1} \cite{5b}. 

Thus one would need transversely polarized gluons to produce one or more mesons. 
In this case the power in $Q$ of the graph would be $Q^0$ and still leading. 
However, the gluon connecting H to B cannot be 
transversely polarized because of helicity conservation of the quarks 
attaching to H. The emission of a  single, transverse gluon would change the helicity
of the involved quark within H where the virtualities are large and helicity 
conservation is valid. Thus such an emission is forbidden and leads to an additional 
suppression by a power of $Q$ bringing down the power in $Q$ to $Q^{-1}$. 
Furthermore, the soft gluon would also have to have transverse polarization, 
which means that the operator for such a three parton state is of higher twist. 
This leads to an additional suppression by a factor of $Q$
showing that the types of graphs in Fig.\ \ref{dijet} truly do not contribute to
the leading power in $Q$.

\subsection{Subtractions, Taylor expansion and gauge invariance.}
\label{together}

The arguments given in Sec.\ VI of \cite{cfs} about subtractions to avoid 
double counting in the leading regions 
carry through without alterations since they are very general in nature
without relying on a particular final state.

The Taylor expansion of the amplitude in terms of the $+$ momentum of the 
collinear-to-A and $-$ momentum of the collinear-to-B partons given in 
SEC.\ VIIB and C of \cite{cfs} is also unaltered, since all
the arguments presented there are independent of the particulars of the 
final meson state.

As far as gauge invariance is concerned, the arguments presented in Sec.\ VIID
of \cite{cfs} 
dealing with the operators appearing in the definition of the skewed 
parton distributions are again unaltered since they rely on general quantum 
field theoretic arguments as presented in \cite{col}, independent of the final 
state, save its overall polarization.

Turning to the mesonic wavefunction, one immediately realizes that one is 
not dealing with the usual single meson wavefunction
anymore but rather a multiple meson wavefunction or generalized 
distribution amplitude, first 
introduced in the case of two pion production in \cite{diehl,polya}.
To obtain the appropriate operators for these new objects, 
one can follow the same line of argument as for the skewed 
parton distributions, and finds once
more the same quark and gluon operators. Due to the additional leading region 
discussed in Sec.\ \ref{leadreg}
we also have to discuss the gluon case as compared to the situation in 
\cite{cfs}. The definition of the generalized distribution amplitudes thus are
\bea
V^q(z,\zeta_i,\mu^2,m^2_V) &=& \frac{1}{2N_c}\int^{\infty}_{-\infty}
\frac{dy^+ e^{-izV^{-}y^{+}}}{4\pi}\langle V_1(v_1)..V_N(v_N)| {\bar \psi}(y^+,0,{\bf 0_{\perp}})
\Gamma {\it P} \psi (0)|0\rangle \nonumber\\
V^g(z,\zeta_i,\mu^2,m^2_V)&=& -\frac{1}{2N_c}\int^{\infty}_{-\infty}
\frac{dy^+ e^{-izV^{-}y^{+}}}{4\pi z V^{-}}\langle V_1(v_1)..V_N(v_N)| G^{-}_{\perp}(y^+,0,{\bf 0_{\perp}})
{\it P} G^{- \perp}(0)|0\rangle ,
\label{disdef}
\eea
with $z$ being the momentum fraction entering the generalized distribution
amplitude, $\zeta_i \equiv v^{-}_{i}/V^{-}$, $i=1,..,N$ characterizing the 
distribution of
longitudinal momentum between the mesons, $\mu^2$ being the renormalization/
factorization scale and $\Gamma = \gamma^{-}, \gamma^{-}\gamma_5, 
\gamma^{-}\gamma^{i}\gamma_5$ depending on the meson species produced. 
Note that our definition differs from those in Ref.\ 
\cite{diehl,polya} in the normalization but agree with that given in \cite{cfs}
for a single meson wavefunction. Also note that in contrast to parton 
distributions or single meson wavefunctions, the above generalized distribution 
amplitudes are not constrained by time reversal invariance to be real valued
functions anymore, since $|V_1,...,V_N \rangle$
differs from $\langle V_1,..,V_N|$ only by hermitian conjugation. As was shown in Ref.\ \cite{polya} in
the case of the 2$\pi$ distribution amplitude, the generalized distribution amplitudes
develop an imaginary part above the appropriate N-particle threshold corresponding
to the contribution of on-shell intermediate states. However, this does not pose 
a problem in the factorization proof, but is rather a phenomenological problem as shown 
in \cite{polya}. This is due to the fact that the above arguments are fully general and concern 
the whole amplitude not its real and/or imaginary parts separately. How the uncut amplitude 
relates to the cut one and how the different space- and time-like contributions arise or do not 
arise, is described in the next paragraph. 

Detailing the just said, one has to keep in mind, that we are dealing with amplitudes, 
thus one would expect a time-ordering of the
operators. This however is not necessary as shown in \cite{diehl1} for regular
distribution amplitudes, skewed and diagonal parton distributions. The basic
observation which helps prove this assertion is that the field operators are
separated by a light-like distance as is also true in the above case. Thus,
after a little bit of fancy footwork, it can be shown that the singularity 
structure in the complex plane of loop and external kinematical momenta of both 
the cut and uncut 
quark-nucleon (gluon-nucleon) scattering amplitude or the 
quark-anti-quark-scattering-into-a-meson amplitude is the same. This proves 
the equality of uncut 
(time-ordered) and cut (not time-ordered) amplitudes. Moreover, the singularity 
structure gives the appropriate support interval for the distributions. 
In the case of the generalized distribution amplitudes, the singularity 
structure is independent on the longitudinal momentum fractions of the 
outgoing mesons \cite{6}, thus one reobtains the case
of the ordinary distribution amplitudes and thus the support interval for
the generalized distribution amplitude is $0 \leq z \leq 1$ as in the case
of ordinary distribution amplitudes.   

As far as evolution is concerned, the function evolves according to the usual
Efremov-Radyushkin-Brodsky-LePage evolution equations. This fact can be understood
with the following reasoning:

       In the general case of skewed parton distributions/distribution 
       amplitudes the evolution equations \cite{muell} depend on a skewedness parameter,
       basically, because the partons connecting the hard scattering subgraph
       H to the collinear subgraph A, carry an unequal amount of $+$ momentum, 
       due to the exclusivity of the final state.
       Note that the skewedness parameter is fixed by the exclusive final state
       \cite{6a}.  
       The partonic lines connecting the hard scattering subgraph H to the
       collinear graph B also carry unequal momentum fraction in the
       $-$ direction, however, in this case the skewedness is fixed to be
       maximal, i.e., $1$ since the total $-$ momentum going into B is
       {\it fixed} by the virtual photon's $-$ momentum $q^-$ to be $Q^2/2xp_+$, 
       which forces one parton to have momentum fraction $(1-z)q^-$ if the other 
       one has momentum fraction $zq^-$ in the zero mass limit. Note that this
       is {\it not} true for the partons connecting H to A.

In the case of exclusive $\gamma^* \gamma$ production of hadrons, we will once
again encounter these functions.

\subsection{Completion of Proof}
\label{proofconcl}

Now we can put together all of the above results and obtain Eq.\ (\ref{faceq}),
the generalized form of Eq.\ (3) of Ref.\ \cite{cfs}.

\section{Proof of Factorization for exclusive $\gamma^*\gamma$ 
production of multiple hadrons.}
\label{facprof2}

\subsection{Statement of factorization Theorem}
\label{state1}

The amplitude of exclusive $\gamma^*\gamma$ production of multiple hadrons
has the following factorized form
\beq
M =\sum_{j} \int^{1}_{0} dz H_{j}(z,\mu^2)
V_j(z,\zeta_1,...,\zeta_N,\mu^2,m^2_V) + \mbox{power suppressed terms}.
\label{faceq1}
\eeq

In the following proof, which is a generalization of the well studied 
factorization of $\gamma^* \gamma \rightarrow \pi^0$ \cite{Brod80}, 
we will follow the same route which we 
outlined in Sec.\ \ref{genproof}. The conventions for the momenta of the 
particles involved in the process are the following in light-cone 
coordinates \cite{diehl}
\begin{eqnarray}
   q'^{\mu } &=& \left ( 0,\frac {Q^{2}+W^2}{\sqrt{2}Q},
   {\bf 0}_{\perp } \right ),
\nonumber\\
   q^{\mu } &=& \left ( \frac{Q}{\sqrt{2}},-\frac {Q}{\sqrt{2}},
   {\bf 0}_{\perp }
\right ),
\nonumber\\
   P^{\mu } &=& \left (\frac{Q}{\sqrt{2}},\frac {W^2}{\sqrt{2}Q},
   {\bf 0 }_{\perp }\right ),
\nonumber\\
\label{momenta1}
\end{eqnarray}
with $W^2 = P^2$ and $P = \sum_{i=1}^{N} h_{i}$. We also define,
the light cone fractions as $\zeta_i \equiv h_i^+/P^+$ which differs 
from the $\zeta_i$ defined above in that we are now using the $+$ instead of $-$ 
components of the different hadron momenta. We work in the limit of 
$Q^2 \rightarrow \infty$ and $W^2 = \mbox{const.}$ which is different form the limit which 
we considered in the above proof where the limit is 
$Q^2/m_{V}^2 \rightarrow \infty$ with $Q^2/W^2 = \mbox{const.}$.

\subsection{Regions}
\label{reg1}

Following the routine outlined in Sec.\ \ref{region} one obtains the 
generalized reduced graphs which are nothing but a 
crossed version of the ones appearing in the proof of factorization for 
deeply virtual Compton scattering (DVCS) \cite{ca}.

\begin{figure}
\centering
\mbox{\epsfig{file=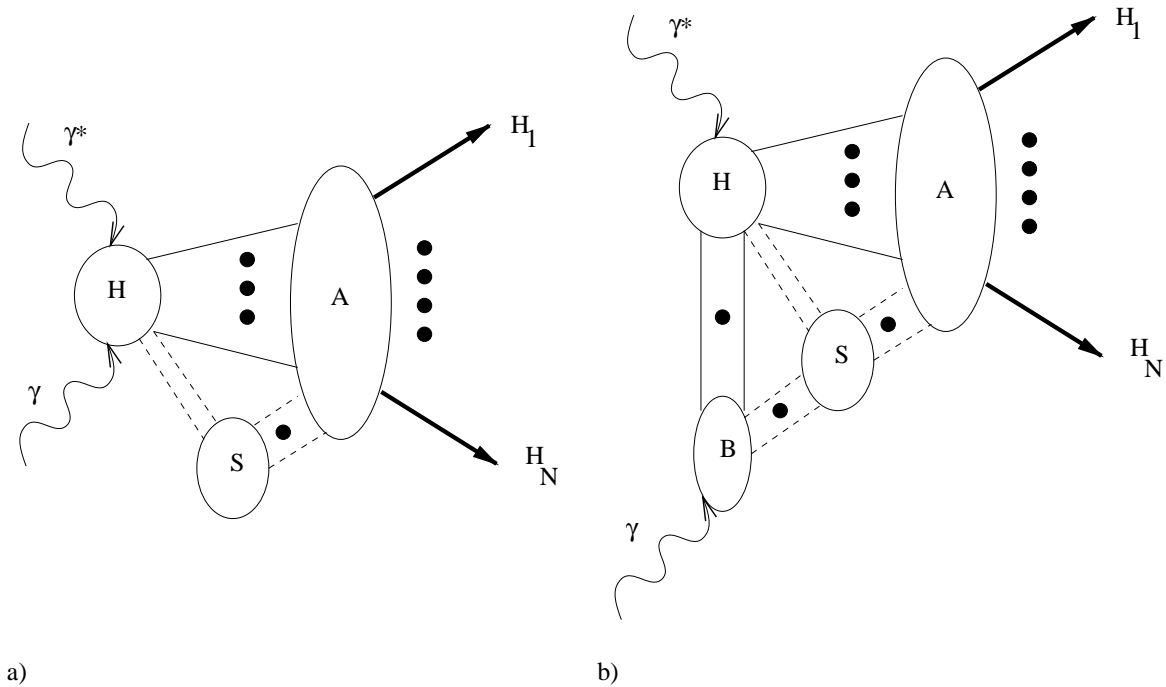,height=9cm}}
\vspace*{5mm}
\caption{Generalized reduced graphs for exclusive $\gamma^*\gamma$ production 
of multiple hadrons.}
\label{genredgr1}
\end{figure}

\subsection{Power Counting}
\label{power1}

In the following proof of a power counting formula, we will closely follow the 
methodology outlined in Ref.\ \cite{cfs}. We find the following power
counting formula analogous to DVCS:
\bea
p(\pi) &=& 4 - n(H) - \#(\mbox{quarks from S to A, B}) - 3\#(\mbox{quarks from
 S to H})\nonumber\\
& & - 2 \#(\mbox{gluons from S to H}).
\label{powerform1}
\eea
where $n(H)$ is now the number of collinear quarks, transversely polarized gluons
and external photons, attaching to the hard part H.
The proof of Eq.\ (\ref{powerform1}) is exactly analogous to the one described in 
Ref.\ \cite{cfs} and also employed in Ref.\ \cite{ca} and will thus not be 
repeated here for the sake of brevity. It is
not too surprising that the powercounting is the same as in DVCS since the
process under consideration is just a crossed version of DVCS.

\subsection{Leading Regions}
\label{leadreg2}

\begin{figure}
\centering
\mbox{\epsfig{file=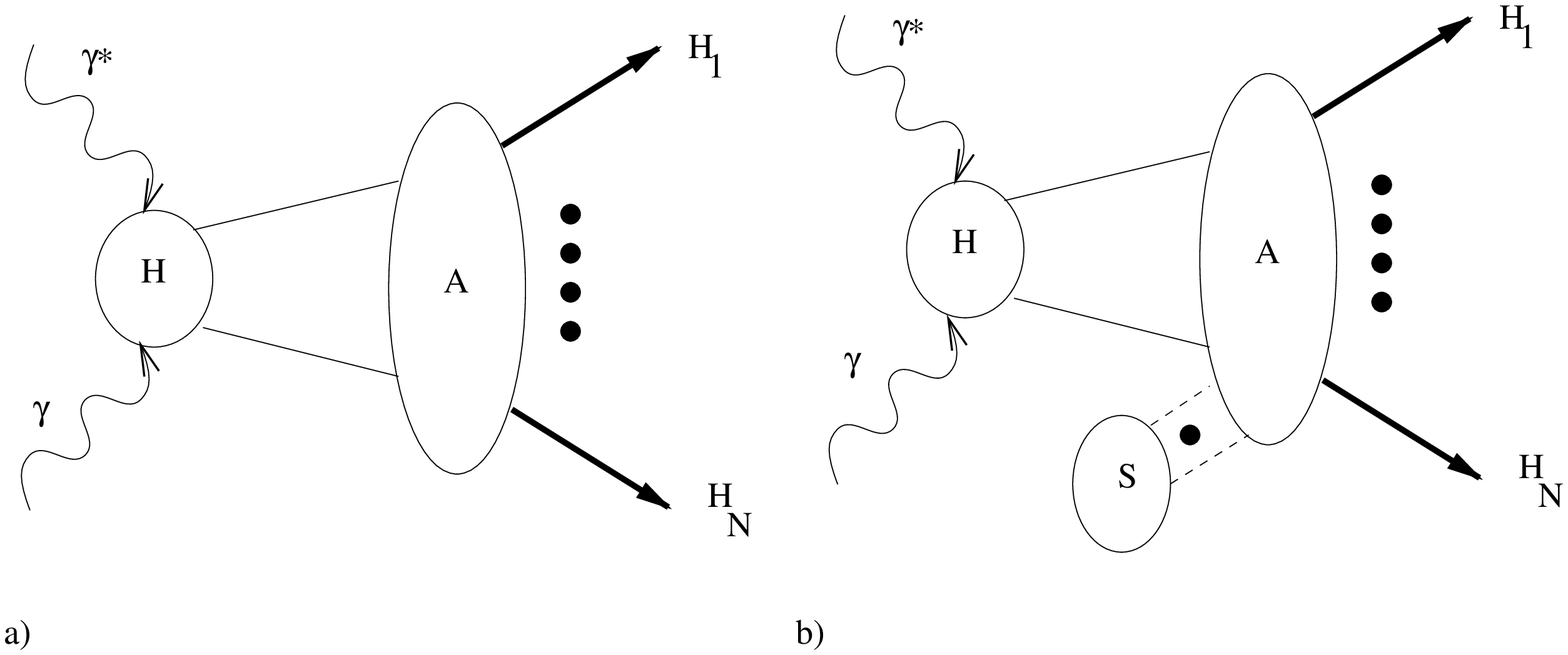,height=6.5cm}}
\vspace*{5mm}
\caption{Leading reduced graphs for exclusive $\gamma^*\gamma$ production 
of multiple hadrons.}
\label{leadredpic}
\end{figure}

The leading regions are found to be those of Fig.\ \ref{leadredpic}, since
they are the ones with the minimal number of particles attaching to H which still
enables the process to proceed. It is not surprising that 
they are again just the crossed DVCS graphs.   

Fig.\ \ref{genredgr1}b \cite{7} is generally suppressed by one power of $Q$ as 
compared to Fig.\ \ref{genredgr1}a and thus does not contribute to the leading
regions, simplifying the proof of factorization tremendously as previously
shown \cite{ca}. Note that this behavior is different from inclusive 
photoproduction where both the resolved and unresolved photon yield the same
power for the cross section. The reason for this not being the case here is 
the following: In inclusive photoproduction there is always only one particle 
attaching to the hard subgraph, either the photon directly in the unresolved 
case or a quark/anti-quark in 
the resolved case, with the other quark/anti-quark going directly into the 
final state making a jet. If the photon is resolved in our case, however, 
both the quark {\it and} anti-quark have to attach to the hard subgraph because of
the exclusiveness of the process.  

Following exactly along the same lines as
outlined in Ref.\ \cite{ca}, one can easily show that there is no soft subgraph
attached to the final state graph A. Hence, Fig.\ \ref{leadredpic}a is the only
leading region in this process and the power of the amplitude for this graph, 
according to our power counting formula Eq.\ (\ref{powerform1}), will be $Q^0$, 
which is the same power as in DVCS. 

\subsection{Subtractions, Taylor expansion and gauge invariance}
\label{stgv1}

The arguments about subtractions to avoid double counting as stated in 
\cite{cfs} carry through without change for the same reasons as mentioned in
Sec.\ \ref{together}.

As far as the Taylor expansion of the amplitude is concerned, this carries 
through in the same way as stated in Ref.\ \cite{ca}, since the structure of
the amplitude in terms of subgraphs is identical to DVCS. There is a 
difference to the case of DVCS, however, which makes the $\gamma^* \gamma$ 
process much easier to deal with. Since the real photon is now in the initial
state, its incoming momentum components are fixed, thus making the hard scattering
function independent of {\it any} non-maximal skewedness parameter \cite{7a}. 
This in turn avoids all the difficulties encountered in DVCS when the skewedness 
parameter was equal to
$x_{Bj}$ \cite{7b}. However, one still has the two 
endpoints of the $z$ integration, $z=0$ and $z=1$ to worry about since at these
endpoints the lines connecting the subgraph A to H become soft and
are thus not collinear-to-A anymore, possibly ruining factorization. Remember 
that the assumption of collinearity of the parton lines connecting A to H
is essential in the proof of factorization. Note, however, that this situation is analogous
to the endpoint case in DVCS \cite{ca}. There it was observed that at the
endpoints, all propagators in the collinear-to-A subgraph have their poles
either above or below the real axis, except one pole that runs off to either $+$ or 
$-$ $\infty$, depending on which endpoint we consider, where it crosses the real axis 
to give a non-vanishing contribution to the amplitude. However this region where the 
pole crosses the real axis is a collinear-to-B region and thus power suppressed.
This in turn means that the distribution amplitude vanishes at those points, thus
restoring the validity of the factorization formula. 

The arguments about gauge invariance as presented in Sec.\ VIID of \cite{cfs} 
carry through
once more without change. The operators encountered are the same as in Sec.\
\ref{together} when dealing with skewed parton distributions and generalized
distribution amplitudes. In this process, the out-state is obviously 
$\langle H(h_1)...H(h_n)|$ 
and the in-state is $|0\rangle $, thus we are dealing with a
generalized distribution amplitude. Since, in this case, we also allow, for 
example, $p\bar p$ final states and have $P$ in $+$ direction, we need to 
slightly change our notation for the generalized distribution amplitudes
\cite{8}
\bea
V^q(z,\zeta_i,\mu^2,m^2_H) &=& \frac{1}{2N_c}\int^{\infty}_{-\infty}
\frac{dy^-}{4\pi}e^{-izP^{+}y^{-}}\langle H(h_1)...H(h_n)| {\bar \psi}(0,y^-,{\bf 0_{\perp}})
\Gamma {\it P} \psi (0)|0\rangle \nonumber\\
V^g(z,\zeta_i,\mu^2,m^2_H)&=& -\frac{1}{2N_c}\int^{\infty}_{-\infty}
\frac{dy^-}{4\pi z P^{+}}e^{-izP^{+}y^{-}}\langle H(h_1)...H(h_n)| G^{+}_{\perp}(0,y^-,{\bf 0_{\perp}})
{\it P} G^{+ \perp}(0)|0\rangle ,
\label{disdef1}
\eea
with  $\Gamma = \gamma^{+}, \gamma^{+}\gamma_5, \gamma^{+}\gamma^{i}\gamma_5$.
The comments made in Sec.\ \ref{together} concerning Eq.\ (\ref{disdef1}) are 
unaltered.

\subsection{Completion of Proof}

After the above said, we just have to assemble the above statements into the
final factorization equation Eq.\ (\ref{faceq1}) to complete the proof.

\section{Conclusions}

We have proved a generalization of the factorization theorem given in 
\cite{cfs} to an arbitrary number of mesons. This generalization led us to 
introduce new nonperturbative objects called generalized distribution 
amplitudes. Furthermore, we proved a factorization theorem for
$\gamma^* \gamma$ production of several hadrons to all orders in perturbation
theory up to power suppressed terms.

\section*{Acknowledgements}

It is a pleasure to acknowledge the more than helpful discussions with
 Markus Diehl, Thierry Gousset and Mark Strikman. I would also like to thank John Collins and Markus Diehl for careful readings of the manuscript. 
This work was supported by E.\ U.\ contract $\#$FMRX-CT98-0194.


\begin{references}
\bibitem{4} The number produced is, of course, limited by the quantum numbers 
            of the initial state.
\bibitem{5}  We define a vector in light cone coordinates by:
		\begin{displaymath}
		V^{\mu }=\left (V^{+},V^{-},V_{\perp } \right )=
		\left ( \frac {V^{0}+V^{3}}{\sqrt {2}},
		\frac {V^{0}-V^{3}}{\sqrt {2}},V^{1,2}\right ).
		\end{displaymath}
\bibitem{5b} The exact power in $Q$ of the type of graphs in Fig.\ \ref{dijet}
             would be $Q^1$ and $Q^3$
             if the gluons connecting A to H and B to H were all scalar!
\bibitem{6}  This can be seen by realizing that the quark-anti-quark loop only 
	     depends on the quark and anti-quark momenta entering the 
	     intermediate state with total momentum $V$ and thus the
             singluarities will not depend on the individual momenta of the 
             mesons but rather the combined momentum $V$. This is also the
             reason why the conjugate variable to $y^+$ is $zV^-$ and not any other
             combination of the $\zeta_iV^-$.
\bibitem{6a} For example, the mass shell condition in deeply virtual Compton
             scattering or exclusive vector meson production and the fact that
             we chose the $-$ component of the produced mesons to be large
             determines the skewedness parameter to be equal to $x$ in the
             above two cases. Note however, that for other final states the
             skewedness parameter is different from $x$.
\bibitem{8}  This is merely a matter of convenience, rather than necessity. One 
	     could have taken the incoming real photon to travel along the $+$ 
	     direction and then only the labeling of the outgoing particles in 
	     Eq.\ (\ref{disdef}) would change.
\end{references}
\end{document}